\documentclass{jpp}
\usepackage{subeqn}
\usepackage{epsfig}

\let\realverbatim\verbatim
\let\realendverbatim\endverbatim
\renewenvironment{verbatim}
  {\par\addvspace{6pt plus 1pt minus 2pt}\realverbatim}
  {\realendverbatim\addvspace{6pt plus 1pt minus 2pt}}

\ifprodtf \else
  \checkfont{eurm10}
  \iffontfound
    \IfFileExists{upmath.sty}
      {\typeout{^^JFound AMS Euler Roman fonts on the system,
                   using the 'upmath' package.^^J}%
       \usepackage{upmath}}
      {\typeout{^^JFound AMS Euler Roman fonts on the system, but you
                   don't seem to have the}%
       \typeout{'upmath' package installed. JPP.cls can take advantage
                 of these fonts,^^Jif you use 'upmath' package.^^J}%
       \providecommand\mathup[1]{##1}%
       \providecommand\mathbup[1]{##1}%
       \providecommand\umu{\umu}%
      }
  \else
    \providecommand\mathup[1]{#1}%
    \providecommand\mathbup[1]{#1}%
    \providecommand\umu{\mu}%
  \fi
\fi

\ifprodtf \else
  \checkfont{msam10}
  \iffontfound
    \IfFileExists{amssymb.sty}
      {\typeout{^^JFound AMS Symbol fonts on the system, using the
                'amssymb' package.^^J}%
       \usepackage{amssymb}%
         
       \let\ge=\geqslant  
      }{}
  \else
  \fi
\fi

\ifprodtf \else
  \IfFileExists{amsbsy.sty}
    {\typeout{^^JFound the 'amsbsy' package on the system, using it.^^J}%
     \usepackage{amsbsy}}
    {\providecommand\boldsymbol[1]{\mbox{\boldmath $##1$}}}
\fi

\newcommand{\eqsto}[2]{Eqs. (\ref{#1}) to (\ref{#2})}
\newcommand{\Eq}[1]{Equation (\ref{#1})}
\newcommand{\sig}{\sigma_{\rm ex}}
\newcommand{\be}{\begin{equation}}
\newcommand{\ee}{\end{equation}}

\newcommand{\eqa}{\begin{eqnarray}}
\newcommand{\eqe}{\end{eqnarray*}}
\newcommand{\eqnu}{\begin{eqnarray}}
\newcommand{\eqne}{\end{eqnarray}}

\newcommand{\eqs}[2]{Eqs. (\ref{#1}) \& (\ref{#2})}

\newcommand{\eq}[1]{Eq. (\ref{#1})}

\newcommand{\eeq}{\end{eqnarray}}

\newdefinition{definition}[theorem]{Definition}

\title[Journal of Plasma Physics]
{Dynamics of Alfv\'en waves in partially ionized astrophysical plasmas}

\author[D. Shaikh]
{D\ls A\ls S\ls T\ls G\ls E\ls E\ls R \ns S\ls H\ls A\ls I\ls K\ls H\ls}
\affiliation{Department of Physics and \\
Center for Space Physics and Aeronomic Research (CSPAR),\\
University of Alabama at Huntsville, Huntsville, AL 35805. USA.\\
{\tt Email:dastgeer.shaikh@uah.edu}}
\date{Nov 7 2009, Revised  Nov 17, 2009}
\pubyear{2009} 
\volume{00}
\part{0}
\pagerange{\pageref{firstpage}--\pageref{lastpage}}
\doi{S0963548301004989}

\begin{document}

\label{firstpage}
\maketitle

\begin{abstract}
We develop a two dimensional, self-consistent, compressible fluid
model to study evolution of Alfvenic modes in partially ionized
astrophysical and space plasmas. The partially ionized plasma consists
mainly of electrons, ions and significant neutral atoms. The nonlinear
interactions amongst these species take place predominantly through
direct collision or charge exchange processes. Our model uniquely
describe the interaction processes between two distinctly evolving
fluids. In our model, the electrons and ions are described by a single
fluid compressible magnetohydrodynamic (MHD) model and are coupled
self-consistently to the neutral fluid via compressible hydrodynamic
equations. Both plasma and neutral fluids are treated with different
energy equations that adequately enable us to monitor non adiabatic
and thermal energy exchange processes between these two distinct
fluids. Based on our self-consistent model, we find that the
propagation speed of Alfvenic modes in space and astrophysical plasma
is slowed down because these waves are damped predominantly due to
direct collisions with the neutral atoms. Consequently, energy
transfer takes place between plasma and neutral fluids. We describe
the mode coupling processes that lead to the energy transfer between
the plasma and neutral and corresponding spectral features.
\end{abstract}


\section{Introduction}
Alfv\'en waves are ubiquitous in laboratory (\cite{lab}), space
(\cite{Gekelman,dastgeer3,dastgeer4,dastgeer5,dastgeer6,dastgeer7})
and astrophysical plasmas (\cite{balsara,shukla2006,shukla2000}).
These waves are essentially electromagnetic fluctuations in magnetized
plasmas that propagate predominantly along an ambient or guide
constant magnetic field. In the context of partially ionized space and
astrophysical plasmas, these waves interact with the neutral gas and
govern numerous properties. For instance, Kulsrud \& Pearce (1969)
noted that the interaction of a neutral gas and plasma can damp
Alfv\'en waves.  Neutrals interacting with plasma via a relative drag
process results in ambipolar diffusion. Ambipolar diffusion plays a
crucial role in the dynamical evolution of the near solar atmosphere,
interstellar medium, and molecular clouds and star formation.
Oishi \& Mac Low (2006) investigated the role of ambipolar diffusion
to set a characteristic mass scale in molecular clouds and find that
substantial structure persists below the ambipolar diffusion scale
because of the propagation of compressive slow mode MHD waves at
smaller scales. Leake et al (2005) showed that the lower chromosphere
contains neutral atoms, the existence of which greatly increases the
efficiency of wave damping due to collisional friction momentum
transfer.  They find that Alfv\'en waves with frequencies above 0.6Hz
are completely damped and frequencies below 0.01 Hz are
unaffected. They undertook a quantitative comparative study of the
efficiency of the role of (ion-neutral) collisional friction, viscous
and thermal conductivity mechanisms in damping MHD waves in different
parts of the solar atmosphere.  It was pointed out by the authors that
a correct description of MHD wave damping requires the consideration
of all energy dissipation mechanisms through the inclusion of the
appropriate terms in the generalized Ohm’s law, the momentum, energy
and induction equations.  Padoan et al (2000) calculated frictional
heating by ion-neutral (or ambipolar) drift in turbulent magnetized
molecular clouds and showed that the ambipolar heating rate per unit
volume depends on field strength for constant rms Mach number of the
flow, and on the Alfv\'enic Mach number.

The neutral gas not only collisionally interacts with plasma, but it
also charge exchanges with plasma protons in many space plasma
applications (\cite{zank1999,dastgeer}). For example, in the local
interstellar medium (LISM), the low density plasma and neutral H gas
are coupled primarily through the process of charge exchange. On
sufficiently large temporal and spatial scales, a partially ionized
plasma is typically regarded as equilibrated; this is the case for the
LISM.

On smaller scales, the edge region of a tokamak (a donut shaped
toroidal experimental device designed to achieve thermonuclear fusion
reaction in an extremely hot plasma) is partially ionized. In a
tokmak, the neutral particles result from effects such as gas puffing,
impurity injection, recombination, charge exchange, and possibly
neutral beam injection processes. The presence of neutrals can
potentially alter the dynamics of zonal flows and cross-field
diffusivity (\cite{singh2004,groebner2001}).

It is worth mentioning that the charge exchange process in
its simpler (and leading order) form can be treated like a friction or
viscous drag term in the fluid momentum equation, describing the
relative difference in the ion and neutral fluid velocities. The drag
imparted in this manner by a collision between ion and neutral also
causes ambipolar diffusion, a mechanism used to describe the Alfv\'en
wave damping by cosmic rays (Kulsrud \& Pearce 1969) and also
discussed by Oishi \& Mac Low (2006) in the context of molecular
clouds.

In this paper we focuse on understanding the propagation
characteristic of Alfv\'en waves by considering the collisional and
charge exchange interactions.  Our model includes both the
interactions simultaneously. In Section 2, we discuss the equations of
a coupled plasma-neutral model, their validity, the underlying
assumptions and the normalizations.  Section 3 describes the results
of our nonlinear, coupled, self-consistent fluid simulations. We find
that propagation characteristics of Alfv\'en waves depends critically
on collision and charge exchange processes. Finally, a summary is
presented in section 4.

\section{MHD Model Equations}
We assume that fluctuations in the plasma and neutral fluids are
isotropic, homogeneous, thermally equilibrated and turbulent. A mean
or constant magnetic field is present. Local mean flows may
subsequently be generated by self-consistently excited nonlinear
instabilities. The boundary conditions are periodic, essentially a box
of plasma.  Most of these assumptions are appropriate to realistic
space and astrophysical turbulent flows. They allow us to use MHD and
hydrodynamic descriptions for the plasma and the neutral components
respectively.  In the context of LISM and outer heliospheric plasmas,
the plasma and neutral fluid remain close to thermal equilibirium and
behave as Maxwellian fluids.  Our model simulates the plasma-neutral
fluid that is coupled via collisions and charge exchange in space (and
astrophysical) plasmas.  The fluid model describing nonlinear
turbulent processes in the interstellar medium, in the presence of
charge exchange, can be cast into plasma density ($\rho_p$), velocity
(${\bf U}_p$), magnetic field (${\bf B}$), pressure ($P_p$) components
according to the conservative form

\be
\label{mhd}
 \frac{\partial {\bf F}_p}{\partial t} + \nabla \cdot {\bf Q}_p={\cal Q}_{p,n},
\ee
where,
\[{\bf F}_p=
\left[ 
\begin{array}{c}
\rho_p  \\
\rho_p {\bf U}_p  \\
{\bf B} \\
e_p
  \end{array}
\right], 
{\bf Q}_p=
\left[ 
\begin{array}{c}
\rho_p {\bf U}_p  \\
\rho_p {\bf U}_p {\bf U}_p+ \frac{P_p}{\gamma-1}+\frac{B^2}{8\pi}-{\bf B}{\bf B} \\
{\bf U}_p{\bf B} -{\bf B}{\bf U}_p\\
e_p{\bf U}_p
-{\bf B}({\bf U}_p \cdot {\bf B})
  \end{array}
\right],\\
{\cal Q}_{p,n}=
\left[ 
\begin{array}{c}
0  \\
{\bf Q}_M + {\bf{F}}_{p,n}   \\
0 \\
Q_E + {\bf U}_p \cdot {\bf{F}}_{p,n}
  \end{array}
\right]
\] 
and
\[ e_p=\frac{1}{2}\rho_p U_p^2 + \frac{P_p}{\gamma-1}+\frac{B^2}{8\pi}.\]
The above set of plasma equations is supplimented by $\nabla \cdot {\bf
B}=0$ and is coupled self-consistently to the  neutral density
($\rho_n$), velocity (${\bf V}_n$) and pressure ($P_n$) through a set
of hydrodynamic fluid equations,
\be
\label{hd}
 \frac{\partial {\bf F}_n}{\partial t} + \nabla \cdot {\bf Q}_n={\cal Q}_{n,p},
\ee
where,
\[{\bf F}_n=
\left[ 
\begin{array}{c}
\rho_n  \\
\rho_n {\bf V}_n  \\
e_n
  \end{array}
\right], 
{\bf Q}_n=
\left[ 
\begin{array}{c}
\rho_n {\bf V}_n  \\
\rho_n {\bf V}_n {\bf V}_n+ \frac{P_n}{\gamma-1} \\
e_n{\bf V}_n
  \end{array}
\right],\\
{\cal Q}_{n,p}=
\left[ 
\begin{array}{c}
0  \\
{\bf Q}_M + {\bf{F}}_{n,p}   \\
Q_E + {\bf V}_n \cdot {\bf{F}}_{n,p}
  \end{array}
\right],
\] 
\[e_n= \frac{1}{2}\rho_n V_n^2 + \frac{P_n}{\gamma-1}.\]

Equations (\ref{mhd}) to (\ref{hd}) form an entirely self-consistent
description of the coupled  plasma-neutral turbulent fluid.

Several points are worth noting.  The charge-exchange momentum sources
in the plasma and the neutral fluids, i.e. Eqs. (\ref{mhd}) and
(\ref{hd}), are described respectively by terms ${\bf Q}_M({\bf
  U}_p,{\bf V}_n,\rho_p, \rho_n, T_n, T_p)$ and ${\bf Q}_M({\bf
  V}_n,{\bf U}_p,\rho_p, \rho_n, T_n, T_p)$. A swapping of the plasma
and the neutral fluid velocities in this representation corresponds,
for instance, to momentum changes (i.e. gain or loss) in the plasma
fluid as a result of charge exchange with the  neutral atoms
(i.e. ${\bf Q}_M({\bf U}_p,{\bf V}_n,\rho_p, \rho_n, T_n, T_p)$ in
Eq. (\ref{mhd})). Similarly, momentum change in the neutral fluid by
virtue of charge exchange with the plasma ions is indicated by ${\bf
  Q}_M({\bf V}_n,{\bf U}_p,\rho_p, \rho_n, T_n, T_p)$ in
Eq. (\ref{hd}). ${\bf{F}}_{p,n} = \nu \rho_p \rho_n ({\bf U}_p-{\bf
  V}_n) = - {\bf{F}}_{n,p}$ is ion-neutral collision force. This force
is self-consistently calculated in our model. In the
absence of charge exchange interactions, the plasma and the neutral
fluid are de-coupled trivially and behave as ideal fluids.  While the
charge-exchange interactions modify the momentum and the energy of
plasma and the neutral fluids, they conserve density in both the
fluids (since we neglect photoionization and
recombination). Nonetheless, the volume integrated energy and the
density of the entire coupled system will remain conserved in a
statistical manner. The conservation processes can however be altered
dramatically in the presence of any external forces. These can include
large-scale random driving of turbulence due to any external forces or
instabilities, supernova explosions, stellar winds, etc. Finally, the
magnetic field evolution is governed by the usual induction equation,
i.e. Eq. (\ref{mhd}), that obeys the frozen-in-field theorem unless
some nonlinear dissipative mechanism introduces small-scale damping.

The underlying coupled fluid model can be non-dimensionalized
straightforwardly using a typical scale-length ($\ell_0$), density
($\rho_0$) and velocity ($v_0$). The normalized plasma density,
velocity, energy and the magnetic field are respectively;
$\bar{\rho}_p = \rho_p/\rho_0, \bar{\bf U}_p={\bf U}_p/v_0,
\bar{P}_p=P_p/\rho_0v_0^2, \bar{\bf B}={\bf B}/v_0\sqrt{\rho_0}$. The
corresponding neutral fluid quantities are $\bar{\rho}_n =
\rho_n/\rho_0, \bar{\bf U}_n={\bf U}_n/v_0,
\bar{P}_n=P_n/\rho_0v_0^2$. The momentum and the energy
charge-exchange terms, in the normalized form, are respectively
$\bar{\bf Q}_m={\bf Q}_m \ell_0/\rho_0v_0^2, \bar{Q}_e=Q_e
\ell_0/\rho_0v_0^3$. The non-dimensional temporal and spatial
length-scales are $\bar{t}=tv_0/\ell_0, \bar{\bf x}={\bf
  x}/\ell_0$. Note that we have removed bars from the set of
normalized coupled model equations (\ref{mhd}) \& (\ref{hd}).  The
charge-exchange cross-section parameter ($\sigma$), which does not
appear directly in the above set of equations (see the subsequent
section for more detail), is normalized as $\bar{\sigma}=n_0 \ell_0
\sigma$, where the factor $n_0\ell_0$ has dimension of (area)$^{-1}$.
By defining $n_0, \ell_0$ through $\sigma_{ce}=1/n_0\ell_0=k_{ce}^2$,
we see that there exists a charge exchange mode ($k_{ce}$) associated
with the coupled plasma-neutral turbulent system.  For a
characteristic density, this corresponds physically to an area defined
by the charge exchange mode being equal to (mpf)$^2$.  Thus the larger
the area, the higher is the probability of charge exchange between
plasma ions and neutral atoms.  Therefore, the probability that charge
exchange can directly modify those modes satisfying $k<k_{ce}$ is high
compared to modes satisfying $k>k_{ce}$.  Since the charge exchange
length-scales are much smaller than the turbulent correlation scales,
this further allows many turbulent interactions amongst the nonlinear
turbulent modes before they undergo at least one charge exchanges.  An
exact quantitive form of sources due to charge exchange in our model
is taken from Shaikh \& Zank (2008) [see also appendix A].

\section{Alfv\'en Wave Dynamics}
We have developed a two-dimensional (2D) nonlinear fluid code to
numerically integrate \eqsto{mhd}{hd}. The 2D simulations are not only
computationally less expensive (compared to a fully 3D calculation),
but they offer significantly higher resolution (to compute inertial
range turbulence spectra) even on moderately-sized clusters like our
local Beowulf.  The spatial discretization in our code uses a discrete
Fourier representation of turbulent fluctuations based on a
pseudospectral method, while we use a Runge Kutta 4 method for the
temporal integration. All the fluctuations are initialized
isotropically with random phases and amplitudes in Fourier space.  A
mean constant magnetic field $B_0$ is assumed along the $y$-direction.
Our algorithm ensures conservation of total energy and mean fluid
density per unit time in the absence of charge exchange and external
random forcing. Additionally, $\nabla \cdot {\bf B}=0$ is satisfied at
each time step.  Our code is massively parallelized using Message
Passing Interface (MPI) libraries to facilitate higher resolution. The
initial isotropic turbulent spectrum of fluctuations is chosen to be
close to $k^{-2}$ with random phases in all three directions.  The
choice of such (or even a flatter than -2) spectrum does not influence
the dynamical evolution as the final state in our simulations
progresses towards fully developed turbulence.

\begin{figure}
\includegraphics[width=14cm]{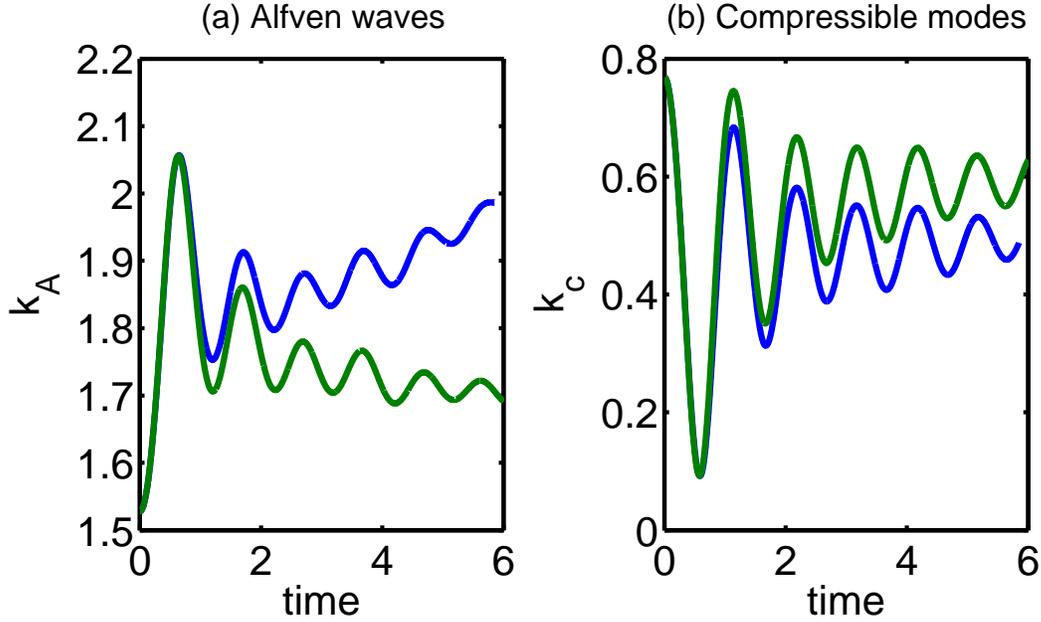}
\caption{Evolution of Alfv\'enic and fast/slow modes in a partially
  ionized plasma. The effect of charge exchange and collisions is
  shown. (a) The propagating Alfv\'en waves damp predominantly due to
  collision (than the charge exchange). (b) Compressive
  (i.e. fast/slow) modes show a less effective tendency towards the
  damping.  The lower curves correspond to a higher strength of
  collisional and charge exchange. The modes and time are measured in
  the normalized units $\ell_0$ and $\ell_0/v_0$ respectively. }
\label{fig1}
\end{figure}

Our goal is to determine the quantative evolution of Alfv\'en waves in
a partially ionized plasma. For this purpose, we distinguish the
Alfv\'enic and non-Alfv\'enic, i.e. corresponding to the compressional
or due to slow and fast magnetosonic modes, contributions to the
turbulent velocity fluctuations.  To identify the distinctive role of
Alfv\'enic and fast/slow (or compressional) MHD modes, we introduce
diagnostics that distinguish the modes corresponding to Alfv\'enic and
slow/fast magnetosonic fluctuations.  Since the Alfv\'enic
fluctuations are transverse, the propagation wave vector is orthogonal
to the velocity field fluctuations i.e.  ${\bf k} \perp {\bf U}$, and
the average spectral energy contained in these (shear Alfv\'enic
modes) fluctuations can be computed as (\cite{dastgeer2})
\[ \langle k_{A} (t) \rangle \simeq \sqrt{\frac{{\sum_{\bf k}}|i {\bf k} \times {\bf U}_{\bf k}|^2}{\sum_{\bf k} |{{\bf U}_{\bf k}}|^2}}.\]
The above relationship leads to a finite spectral contribution from
the $| {\bf k} \times {\bf U}_{\bf k}|$ characteristic turbulent
Alfv\'enic modes.  On the other hand, fast/slow (i.e. compressive)
magnetosonic modes propagate longitudinally along the velocity field
fluctuations, i.e.  ${\bf k} \parallel {\bf U}$ and thus carry a
finite component of energy corresponding {\em only} to the $i {\bf k}
\cdot {\bf U}_{\bf k}$ part of the velocity field, which can be
determined from the following relationship
\[ \langle k_{c} (t) \rangle \simeq \sqrt{\frac{{\sum_{\bf k}}|i {\bf k} \cdot {\bf U}_{\bf k}|^2}{\sum_{\bf k} |{{\bf U}_{\bf k}}|^2}}.\]
The expression of $k_{c}$ essentially describes the modal energy
contained in the non-solenoidal component of the MHD turbulent modes.

\begin{figure}[t]
\includegraphics[width=12cm]{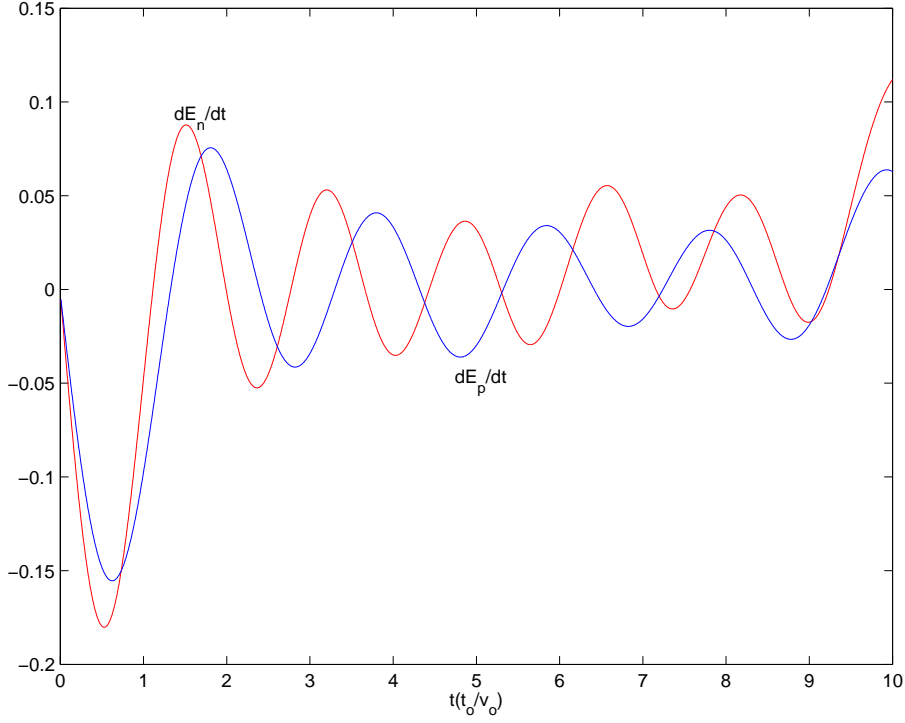}
\caption{The rate of transfer of energy (or energy exchange) between
  the plasma and neutral fluids is shown. Clearly the decay rates of
  energy in one fluid is compensated by the increased rates in the
  other fluid such that entire energy of the coupled plasma-neutral
  system is conserved provided no external source of dissipation
  exists in the system. }
\label{fig2}
\end{figure}

The evolution of Alfv\'enic and compressive modes is shown in Fig (1).
We vary charge exchange strength and collision parameter in our
simulation to examine their effects on the propagation of Alfv\'enic
and compressive modes in the partially ionized plasmas. Clearly, the
two processes operate on different time and length scales and are
self-consistently modeled in our simulations (see Eqs 1 ans 2).  We
find that charge exchange and collisional interactions jointly damp
the Alfv\'en waves. This is shown in Fig (1a) for $256^2$ modes in a
two dimensional box of length $2\pi \times 2\pi$. The other parameters
in our simulations are; charge exchange $k/k_{ce} \sim 0.1 - 0.01$,
fixed time step $dt=10^{-3}$, and collision parameter $\nu \sim 0.1 -
0.001$. The background constant magnetic field $B_0=0.5$. Our
simulations are fully nonlinear because the ratio of the mean and
fluctuating magnetic fields $\delta {\bf B}/{\bf B}_0 \sim 1$.  It is
clear from Fig (1a) that the increasing strength of charge exchange
and collision leads to a damping of the Alfv\'en waves. Similarly, the
compressive modes are also damped by virtue of the increased charge
exchange and collision. This is shown in Fig (2a). We further find
that collisional interactions are more promising to damp the MHD waves
in comparison with the charge exchange process.

It is further interesting to note that rate of change of energy in
plasma fluid is balanced precisely by rate of increase of energy in
the neutral fluid by an equivalent amount. This is shown in Fig (2).
This is true as long as there are no external sources of dissipation
in the two fluids.  We find from Fig (2) that rate of decay of plasma
is followed by the rate of increase of energy in the neutral
fluid. Thus on an average, the net transfer of energy from plasma to
the neutral fluid (or vice versa) is nearly zero. It then turns out
that the collisional and charge exchange processes dampen the wave
activities, while the energy cascade rates in the two fluids continue
to remain constant in the absence of external dissipation. The effect
of a random driving does not change the above results, because the
random froces isotropically drive the plasma and neutral fluids.

\section{Conclusion}
In this work, we develop a two dimensional self-consistent model of
plasma and neutral fluids that are coupled through charge exchange and
collisional interactions. We concentrate on understanding the
propagation of MHD waves in the partially ionized space and/or
astrophysical plasmas.  One of the most important points to emerge
from our studies is that charge exchange in combination with
collisional interactions modify the propagation characteristics of
Alfv\'en and compressive modes. The wave activities are damped.  By
contrast, the energy transfer rates in the plasma and neutral
components continue to remain constant. It is important to note that
the collision and charge exchange processes introduce disparate time
and length scales in the coupled plasma-neutral fluid system. We find
that on scales $\ell\ge\ell_{ce}$, the coupled plasma system evolves
differently than the uncoupled system where large-scale turbulent
fluctuations are strongly correlated with charge-exchange.  We add
finally that our self-consistent model can be useful in studying
turbulent dynamics of partially ionized plasma in the magnetosphere of
Saturn and Jupiter where outgassing from moons and Io and Encephalus
introduces a neutral gas into the plasma (\cite{dastgeer}).

\section{Acknowledgement}
The support of NASA(NNG-05GH38) and NSF (ATM-0317509) grants is  acknowledged.

\appendix
\section{Charge Exchange Sources}
The charge exchange terms can be obtained from the Boltzmann transport
equation that describes the evolution of a neutral distribution
function $f_n=f({\bf x}, v_x, v_y, v_z, t)$ in a six-dimensional phase
space defined respectively by position and velocity vectors $({\bf x},
v_x, v_y, v_z)$ at each time $t$. Here we follow Pauls et al. (1995)
in computing the charge exchange terms from various moments of the
Boltzmann equation. The Boltzmann equation for the neutral
distribution contains a source term proportional to the proton
distribution function $f_p$ and a loss term proportional to the
neutral distribution function $f_n$,
\be
\label{bol}
\frac{\partial f_n}{\partial t} + {\bf v}_n \cdot \nabla f_n +
\frac{\bf F}{m} \cdot \nabla_{{\bf v}_n} f_n = P - L 
\ee
\[P=f_{p}({\bf x}, v_x, v_y, v_z, t) \int f_n({\bf x}, v_x, v_y, v_z, t) |{ v}_n- {u}_p|
\sig(v_{rel}) d^3{v}_n \]
\[L=f_n({\bf x}, v_x, v_y, v_z, t) \int
f_p({\bf x}, v_x, v_y, v_z, t) |{ u}_p- { v}_n| \sig(v_{rel})
d^3{ u}_p.\] 
Here $f_p, ~{u}_p$ represent respectively the
proton distribution function and velocity.  $\sig$ is the charge
exchange cross-section (between neutrals and plasma protons), $m$ is
the mass of particle, and ${\bf F}$ represents forces acting on the
fluid.  The charge exchange parameter has a logarithmically weak
dependence on the relative speed ($v_{rel}=|{u}_p- {v}_n|$) of
the neutrals and the protons through $\sig = [(2.1-0.092 \ln
  (v_{rel})) 10^{-7} cm]^2$ [\cite{fite}].  This cross-section is valid
as long as energy does not exceed $1eV$, which usually is the case in
the inner/outer heliosphere. Beyond $1eV$ energy, this cross-section
yields a higher neutral density. This issue is not applicable to our
model and hence we will not consider it here.  The density, momentum,
and energy of the thermally equilibrated Maxwellian proton and neutral
fluids can be computed from \eq{bol} by using the zeroth, first and
second moments $\int f_{\xi} d^3\xi, \int m{\bf \xi} f_{\xi} d^3\xi$
and $\int m\xi^2/2 f_{\xi} d^3\xi$ respectively, where $\xi={u}_p$
or ${v}_n$. Since charge exchange conserves the density of the
proton and neutral fluids, there are no sources in the corresponding
continuity equations. We, therefore, need not compute the zeroth
moment of the distribution function.  Computing directly the first
moment from \eq{bol}, we obtain the neutral fluid momentum equation as
given by \eq{hd}.  The entire rhs of \eq{bol} can now be replaced by a
momentum transfer function ${\bf Q}_M({v}_n,{u}_p) $ which
reads 
\be
\label{qm}
 {\bf Q}_M({v}_n,{u}_p) = \bar{\mu}({u}_p,{v}_n)-\bar{\mu}({v}_n,{u}_p),
\ee
where ${\bf Q}_M$ and $ \bar{\mu}$, the transfer integral, are 
vector quantities. The transfer integrals describe  the
charge exchange transfer of momentum from proton to neutral fluid and
vice versa.  The first term on the rhs of
\eq{bol} can be expressed by
\[\bar{\mu}({u}_p,{v}_n) = f_p({\bf x}, v_x,v_y,v_z,t)\beta({u}_p,{v}_n)\]
where,
\[\beta({u}_p,{v}_n)=\int f_n({\bf x}, v_x,v_y,v_z,t) |{v}_n- {u}_p|
\sig(v_{rel}) d^3{v}_n.\]
Considering a Maxwellian distribution for the neutral atoms, we simplify 
$\beta({u}_p,{v}_n)$ as
follows,
\[\beta({u}_p,{v}_n)= \sig n_n V_{T_n} 
\sqrt{\frac{4}{\pi}+\frac{({v}_n- {u}_p)^2}{V_{T_n}^2}}.\] Note that
the above expression emerges directly from a straightforward
integration of sources in the rhs of the Boltzmann \eq{bol}. To obtain
the expression for the momentum transferred from proton to neutral (or
vice versa), we need to take a second moment of the
$\bar{\mu}({u}_p,{v}_n)$ expression. This is shown in the following,
\[\bar{\mu}({u}_p,{v}_n) = m {\bf v}_n I_0({u}_p,{v}_n) + m ({\bf u}_p-{\bf v}_n) I_1({u}_p,{v}_n).\]
where $I_0$ and $I_1$ are transfer integrals that can be written as follows,
\[I_0({u}_p,{v}_n)= \int  f_p({\bf x}, v_x,v_y,v_z,t)\beta({u}_p,{v}_n) ~d^3{u}_p;\]
\[I_1({u}_p,{v}_n)= \int {v}_n f_p({\bf x}, v_x,v_y,v_z,t)\beta({u}_p,{v}_n) ~d^3{u}_p.\]
Assuming a Maxwellian distribution for plasma protons and using
$\beta({u}_p,{v}_n)$ from the above expression, we can
straightforwardly evaluate the transfer integrals $I_0$ and $I_1$ (see
Pauls et al (1995) for details).  We further write the
complete form of the first term on the rhs of Eq. (4) as follows,
\[\bar{\mu}({u}_p,{v}_n) = m\sig n_p n_n \left[U_{{u}_p,{v}_n}^\ast{\bf v}_n - ({\bf u}_p-{\bf v}_n)
\frac{  V_{T_n}^2}{\delta V_{{u}_p,{v}_n}} \right].\] In a similar
manner, we can evaluate the second term on the rhs of \eq{bol}, which
yields the following form,
\[\bar{\mu}({v}_n,{u}_p) = m\sig n_{p}n_n \left[U_{{v}_n,{u}_p}^\ast {\bf u}_p - ({\bf v}_n-{\bf u}_p) \frac{V_{T_p}^2}{\delta V_{{v}_n,{u}_p}} \right],\]
where 
\[\delta V_{{u}_p,{v}_n} =
  \left[4\left(\frac{4}{\pi}V_{T_p}^2+\Delta U^2 \right)
  +\frac{9\pi}{4}V_{T_n}^2 \right]^{1/2},\]
\[ \delta V_{{v}_n,{u}_p} = \left[4\left(\frac{4}{\pi}V_{T_n}^2+\Delta U^2 \right)
  +\frac{9\pi}{4}V_{T_{n}}^2 \right]^{1/2}\]
\[U^\ast=U_{{u}_p,{v}_n}^\ast = U_{{v}_n,{u}_p}^\ast =
  \left[\frac{4}{\pi}V_{T_{p}}^2+\frac{4}{\pi}V_{T_{n}}^2 +\Delta
  U^2\right]^{1/2}, \Delta {u} = {u}_p-{v}_n.\]  
On
  substituting these expressions in the momentum transfer function, we
  obtain 
\be
\label{qm2}
 {\bf Q}_M({v}_n,{u}_p) = m\sig n_{p}n_n ({\bf v}_n -{\bf
   u}_p) \left[ U^\ast + \frac{V_{T_n}^2}{\delta V_{{u}_p,{v}_n}} -\frac{V_{T_p}^2}{\delta V_{{v}_n,{u}_p}}
   \right].  
\ee 
\Eq{qm2} together with \eq{hd} yields the momentum equation for
the neutral gas. Swapping the plasma and neutral fluid velocities
yields the corresponding source term for the proton fluid momentum
equation.  The gain or the loss in neutral or proton fluid momentum
depends upon the charge exchange sources, which depend upon the
relative speeds between neutrals and the protons. The thermal speeds
of proton and neutral gas in \eq{qm2} are given respectively by $
V_{T_p}^2 = 2K_BT_{p}/m$ (the factor 2 arises because of thermal
equilibration in that it is assumed that the temperature of the plasma
electrons and protons are nearly identical so that $T_p = T_e + T_{\rm
  proton} \simeq 2T_p$) and $V_{T_n}^2 = K_BT_{n}/m$. The
corresponding temperatures are related to the pressures by $
P_{p}=2n_{p}K_BT_{p}$ and $P_{n}=n_{n}K_BT_{n}$, where $n_n, T_n,
n_{p}, T_{p}$ are respectively the neutral and plasma density and the
temperature, and $K_B$ is the Boltzmann constant.

The  moment, $\int m\xi^2/2 f_{\xi} d^3\xi$, of the Boltzmann
\eq{bol} yields an energy equation for the neutral fluid whose
rhs contains the charge exchange energy transfer function 
\[Q_E({v}_n,{u}_p) = \eta({u}_p,{v}_n)-\eta({v}_n,{u}_p),\]
where $\eta({u}_p,{v}_n), ~\eta({v}_n,{u}_p)$ are the
energy transfer (from neutral to proton and vice versa) rates. These
functions can be computed as follows: 
\[\eta({u}_p,{v}_n) = \frac{1}{2}
mV_n^2 \sig n_{p}n_n U^\ast + \frac{3}{4}m V_{T_n}^2 \sig n_{p}n_n \Delta
V_{{u}_p,{v}_n}- m \sig n_pn_n {\bf v}_n\cdot ({\bf u}_p-{\bf
v}_n) \frac{V_{T_n}^2}{\delta V_{{u}_p,{v}_n}}\]
 and 
\[\eta({v}_n,{u}_p) = \frac{1}{2}mV_{n}^2 \sigma n_{p}n_n U^\ast +  \frac{3}{4}m
V_{T_{p}}^2 \sig n_{p}n_n \Delta V_{{v}_n,{u}_p}- m \sig
n_{p}n_n {\bf u}_p \cdot ({\bf v}_n-{\bf u}_p) \frac{V_{T_{p}^2}}{\delta
V_{{v}_n,{u}_p}}.\]  
The total energy transfer from neutral to proton fluid, due to charge
exchange, can then be written as,
\eqa
\label{qe}
Q_E({v}_n,{u}_p) &=& \frac{1}{2}m \sig  n_{p}n_n U^\ast (V_n^2-U_p^2)+
\frac{3}{4}m  \sig  n_{p}n_n (V_{T_{n}}^2\Delta V_{{u}_p,{v}_n}-V_{T_{p}}^2
\Delta V_{{v}_n,{u}_p}) 
 \nonumber \\
&&-m \sig  n_{p}n_n
\left[ {\bf v}_n \cdot ({\bf u}_p-{\bf v}_n)\frac{V_{T_n}^2}{\delta V_{{u}_p,{v}_n}}-
{\bf u}_p \cdot ({\bf v}_n-{\bf u}_p)\frac{V_{T_{p}^2}}{\delta V_{{v}_n,{u}_p}}\right],
\eeq
\[\Delta V_{{u}_p,{v}_n} = 
[\frac{4}{\pi}V_{T_p}^2+\frac{64}{9\pi}V_{T_n}^2 +\Delta U^2]^{1/2},\] 
{\rm and} ~
\[\Delta V_{{v}_n,{u}_p} = [\frac{4}{\pi}V_{T_n}^2+
\frac{64}{9\pi}V_{T_{p}}^2 +\Delta U^2]^{1/2}.\]
A similar expression for the energy transfer charge exchange source
term of plasma energy in \eq{mhd} can be obtained by exchanging the
plasma and neutral fluid velocities. In the normalized momentum and
energy charge exchange source terms, the factor $m\sig$ in
\eqs{qm2}{qe} is simply replaced by $\bar{\sigma}$.




\end{document}